\documentclass{article}

\usepackage[
  paperwidth=8.5in, paperheight=11in,
  left=0.75in, right=0.75in,
  top=0.90in, bottom=0.90in,
  columnsep=0.25in
]{geometry}

\usepackage{amsmath,amssymb,amsthm}
\usepackage{physics}
\usepackage{booktabs}
\usepackage{microtype}
\usepackage{graphicx}
\usepackage{enumitem}
\usepackage{cite}
\usepackage{url}
\usepackage{titlesec}
\usepackage{abstract}
\usepackage{caption}
\usepackage{multirow}
\usepackage{mathtools}

\usepackage[T1]{fontenc}
\usepackage[utf8]{inputenc}
\usepackage{mathptmx}
\usepackage{helvet}

\usepackage{mdframed}
\usepackage{xcolor}

\newmdenv[
  leftline=true, rightline=false, topline=false, bottomline=false,
  linewidth=2pt, linecolor=black!50,
  backgroundcolor=black!4,
  innerleftmargin=8pt, innerrightmargin=6pt,
  innertopmargin=5pt, innerbottommargin=5pt,
  skipabove=6pt, skipbelow=6pt
]{resultbox}

\newcommand{\Jeff}{J_{\mathrm{eff}}}
\newcommand{\Pform}{\Psi_{\mathrm{form}}}

\newcommand{\bTF}{\beta_{\mathrm{TF}}}
\newcommand{\gTF}{\gamma_{\mathrm{TF}}}
\newcommand{\zTF}{z_{\mathrm{TF}}}
\newcommand{\chiTF}{\chi_{\mathrm{TF}}}
\newcommand{\kB}{k_{\mathrm{B}}}
\newcommand{\Order}{\mathcal{O}}

\usepackage[colorlinks=true, citecolor=blue, urlcolor=blue, linkcolor=blue, breaklinks]{hyperref}

\begin{document}

\title{\bfseries Composite-Operator Scaling on Triadic Hypergraphs:\\ Formation Transitions in Multi-Agent Architectures\\ with Three-Body Coupling}

\author{Eduardo Salazar\\ Nebula Technology Lab\\ \texttt{eduardo.salazar@nebulalab.ae}}

\date{}

\maketitle

\begin{abstract}
This paper studies phase transitions on dynamic triadic hypergraphs, in which a continuous formation field $\phi_i$ evolves under stochastic Ginzburg--Landau 
dynamics with an explicit cubic three-body coupling $g_\tau\phi_i\phi_j\phi_k$ sourced by the hypergraph topology. At the same time, a discrete opinion variable 
$s_i\!\in\!\{-1,+1\}$ undergoes Kawasaki exchange governed by a Hamiltonian containing both pairwise alignment and an irreducible three-body energy 
$-\lambda_\tau\!\prod_{a\in\tau}\!s_a$. Near the formation critical point, the cubic coupling is subleading in the reduced temperature and the transition remains 
continuous, controlled at leading order by a pairwise Ising baseline with renormalized coupling $\Jeff = J + \gamma w$. 

The dominant macroscopic observable is the triadic formation correlator $\Psi_{\rm form} \equiv \langle \phi_i \phi_j \phi_k \rangle$. This is a $k=3$ composite 
operator built over the underlying $\mathbb{Z}_2$-symmetric order parameter. Composite-operator scaling then yields the effective exponents $\beta_{\rm TF} = 3/2$ and 
$\gamma_{\rm TF} = -1$. In particular, the susceptibility conjugate to $\Psi_{\rm form}$ \emph{vanishes} at the critical temperature $T_c$ rather than diverging, in 
contrast to the divergence that characterizes scalar (pairwise) order parameters.

The exact partition function of the minimal triad on $\{-1,+1\}^3$ identifies a crossover scale $T^*\!=\!4\Jeff/\ln 3$ above which the composite and primary 
order parameters exchange dominance in the fluctuation spectrum. A field-theoretic two-point function argument reproduces the same vanishing susceptibility. When 
the three-body coupling is restored ($\lambda\neq 0$) the transition becomes first-order, with a critical endpoint at $\lambda=0$. The composite-operator exponent 
relations $\bTF = 3\beta_{\mathrm{Ising}}$ and $\gTF = \gamma_{\mathrm{Ising}} - 4\beta_{\mathrm{Ising}}$ hold exactly in arbitrary dimension, on dense hypergraphs, 
via cluster decomposition, and the vanishing-susceptibility signature persists for $d \geq 3$ but fails in $d=2$. A Mori--Zwanzig projection with multi-timescale 
memory kernel yields a continuously tunable dynamical exponent $\zTF$.

The formation susceptibility emerges as the natural diagnostic observable for triadic interaction networks (the structure underlying neural assemblies and 
other higher-order systems) which this analysis brings within an exactly solvable statistical-mechanics framework.
\end{abstract}

\vspace{0.8em}
\vspace{0.8em}

\section{Introduction}

Phase transitions in networks of interacting units underpin neural computation~\cite{hopfield1982}, social opinion phenomena and collective 
behavior~\cite{castellano2009,galam2008}, and, more recently, coupled systems on higher-order networks with emergent collective behavior~\cite{salazar2026}. 
In every canonical model, be it Ising, Hopfield, voter or Potts, interactions are pairwise. The energy is a sum over dyadic terms.

Biological neural systems and a growing class of engineered systems, by contrast, exhibit \emph{higher-order dependencies} that resist reduction to dyadic 
terms~\cite{buzsaki2006}. The formation of functional neural assemblies and the synchronization of task-specific computational modules all involve coordinated activity 
among three or more units simultaneously~\cite{buzsaki2010}. Capturing such dependencies clearly exceeds a dyadic structure. Even the pairwise-weight Boltzmann machine 
requires the simultaneous statistics of hidden-unit assemblies to represent joint distributions that cannot be encoded by pairwise connections alone~\cite{ackley1985}.

\newpage

The structural claim of this paper is general. In any system on a $k$-uniform hypergraph whose dominant collective observable is a $k$-body correlator 
$O_k\!\equiv\!\langle\phi^k\rangle$ over a $\mathbb{Z}_2$-symmetric mean-field order parameter $m$, composite-operator scaling theory~\cite{goldenfeld2018,stanley1971} 
implies effective exponents
{%
\setlength{\abovedisplayskip}{5pt}%
\setlength{\abovedisplayshortskip}{2pt}%
\setlength{\belowdisplayskip}{5pt}%
\setlength{\belowdisplayshortskip}{3pt}%
\begin{equation}
  \beta_k = \frac{k}{2}, \quad \gamma_k = 2-k, \quad \text{for} \quad k \geq 1.
  \label{eq:general}
\end{equation}%
}%

\noindent The general-$d$ statement is the rigidity relation of Result~8 (see Sec.~\ref{sec:beyondMF}) subject to the locality condition stated there.

These follow from the scaling dimension $\Delta_{O_k} = k\Delta_\phi$ at $d_c=4$, where $\Delta_\phi=1$ is the mean-field scaling dimension of the fluctuating 
field operator. The standard pairwise case ($k\!=\!1$, observable $= m$) recovers $(\beta,\gamma) = (1/2,\,1)$. For $k\!=\!2$, the susceptibility is finite at 
$T_c$ rather than diverging, but for $k\!\geq\!3$ it \emph{vanishes}. The non-diverging, vanishing susceptibility profile is a signature of triadic (or higher) 
observable structure. Within the composite-operator family \eqref{eq:general}, $k\!=\!3$ is the lowest order for which the conjugate susceptibility vanishes 
rather than merely staying finite ($k\!=\!2$) or diverging ($k\!=\!1$), and it is the natural case for systems with triadic interactions.

We derive the $k=3$ case in full. Higher-order interactions on hypergraphs and simplicial complexes have attracted sustained attention in statistical physics, 
from $p$-spin models~\cite{derrida1981}, the retrieval dynamics of modern Hopfield networks~\cite{ramsauer2021} and higher-order associative memories~\cite{krotov2016}, 
to recent work on simplicial synchronization and percolation~\cite{millan2020,bianconi2021}. The triadic case ($k=3$) is the simplest instance that exhibits the 
vanishing-susceptibility regime, and arises naturally in any system whose elementary interaction units are three-body. Our analysis takes as its reference model the 
COGENT$^{\textbf{3}}$ system~\cite{salazar2026}, a stochastic dynamical framework on triadic hypergraphs that builds on~\cite{cirigliano2024}. Its formation field 
$\phi_i \in \mathbb{R}$ evolves under a Ginzburg--Landau potential with cubic three-body coupling $g_\tau\phi_i\phi_j\phi_k$, while a discrete opinion variable 
$s_i \in \{-1,+1\}$ undergoes Kawasaki exchange with both pairwise alignment and an irreducible three-body energy $-\lambda_\tau\prod_{a\in\tau}s_a$. Three structurally 
conserved quantities (formation norm, total opinion, and integrated memory) constrain the dynamics. 

COGENT$^{\textbf{3}}$ provides a representative macroscopic model of the broader class of systems dominated by higher-order interactions. In Sec.~\ref{sec:reduction} we 
show that near the formation critical point, the cubic coupling $g_\tau$ is subleading in the reduced temperature and the critical behavior is controlled at leading 
order by a pairwise Ising baseline with renormalized coupling $\Jeff$.

The goals set for this paper are quite simple: (i)~solve the triad partition function exactly on $\{-1,+1\}^3$; (ii)~derive the thermodynamic-limit transition by 
mean-field theory; (iii)~determine the critical scaling of the formation order parameter $\Pform\equiv\langle\phi_i\phi_j\phi_k\rangle$; (iv)~compute the Mori--Zwanzig 
memory correction to the dynamical critical exponent; (v)~determine the phase diagram when the three-body coupling is restored, identifying the first-order transition 
line and its critical point; and (vi)~establish the composite-operator exponent relations beyond mean field in arbitrary dimension.


\section{The Pairwise Baseline}
\label{sec:triad}

\subsection*{Hamiltonian}
Define $\tau=\{i,j,k\}$ as a triad with binary formation fields $\phi_a\in\{-1,+1\}$ and Hamiltonian given by
\begin{equation}
  H_\tau = -J\!\sum_{\langle ab\rangle\subset\tau}\!\phi_a\phi_b
           - h\sum_{a\in\tau}\phi_a
           + \frac{\gamma}{2}\sum_{\langle ab\rangle\subset\tau}
             w_{ab}(\phi_a-\phi_b)^{2}
  \label{eq:ham}
\end{equation}
\noindent where $J>0$ is the pairwise alignment coupling, $h$ an external symmetry-breaking field, $\gamma>0$ the gradient coefficient, and $w_{ab}=w$ denote uniform 
edge weights.

\subsection*{Exact reduction}
On $\{-1,+1\}$ the following identity
\begin{equation}
  (\phi_a-\phi_b)^2 = 2(1-\phi_a\phi_b)
  \label{eq:identity}
\end{equation}
\noindent holds \emph{exactly}. Applying \eqref{eq:identity} to all three edges of $\tau$, and inserting into \eqref{eq:ham}, yields
\begin{equation}
  H_\tau = -\Jeff\!\sum_{\langle ab\rangle}\!\phi_a\phi_b - h\sum_a\phi_a + E_0
  \label{eq:hamred}
\end{equation}
\noindent with the renormalized coupling and offset
\begin{equation}
  \Jeff = J+\gamma w \quad \text{with} \quad E_0 = 3\gamma w.
  \label{eq:jeff}
\end{equation}

The reduction in \eqref{eq:hamred} is exact for the opinion sector of the Hamiltonian $H_\text{group}$ in the absence of the three-body coupling $\lambda_\tau$. The full 
COGENT$^{\textbf{3}}$ group Hamiltonian (see Eq. 22 of~\cite{salazar2026}) includes an additional term $-\sum_\tau \lambda_\tau \prod_{a \in \tau} s_a$ that does not 
admit pairwise reduction. The pairwise Ising model is therefore the $\lambda_\tau = 0$ limit of the full theory. In the implementation, $\lambda_\tau / J = 1.0$, hence 
the three-body coupling is not perturbatively small. The pairwise model therefore serves as an analytically tractable baseline. Deviations from its predictions quantify 
the effect of the irreducible triadic interaction.

The gradient term does not frustrate the system but cooperates with the alignment interaction, renormalizing $J$ upward by the amount $\gamma w$. The energy is thus 
exactly pairwise, with the triadic physics (within this baseline) entering through the composite observable $\Pform$ rather than the Hamiltonian structure. 
Equation~\eqref{eq:jeff} is \emph{exact} on the binary hypercube: the minimal triad maps exactly onto a renormalized Ising model, enabling analytical solution of the 
baseline mechanics before evaluation of the composite higher-order observables. No approximation is involved in the reduction from~\eqref{eq:ham} to~\eqref{eq:hamred} 
itself.

\subsection*{Exact partition function}
At $h=0$ the eight configurations of $\{-1,+1\}^3$ cluster into two symmetry classes under $\phi\to-\phi$, as shown in Table~\ref{tab:configs}. 

\begin{table}[h]
  \centering
  \caption{Configuration classes of the triad at $h=0$. Degeneracy $g$ counts
    symmetry-equivalent states, with energies that follow from \eqref{eq:hamred}:
    Class~A gives $-(J{+}\gamma w)\cdot3+3\gamma w=-3J$ and Class~B gives 
    $(J{+}\gamma w)+3\gamma w=J+4\gamma w$.}
  \label{tab:configs}
  \medskip
  \begin{tabular*}{0.6\linewidth}{@{\extracolsep{\fill}}lccc@{}}
    \toprule
    Class & Representative & $H$ & $g$ \\
    \midrule
    A (aligned)  & $(+{,}+{,}+)$ & $-3J$         & 2 \\
    B (majority) & $(+{,}+{,}-)$ & $J+4\gamma w$ & 6 \\
    \bottomrule
  \end{tabular*}
\end{table}

\noindent We then obtain the result below.

\begin{resultbox}
  \textbf{Result 1 (Exact partition function)}
  The partition function of the triadic Ising model at $h=0$ is
  \begin{equation}
    \boxed{Z_0 = 2\,e^{3\beta J} + 6\,e^{-\beta(J+4\gamma w)}}
    \label{eq:Z}
  \end{equation}
  where $\beta=(\kB T)^{-1}$. This is \emph{exact} on $\{-1,+1\}^3$.
\end{resultbox}

\subsection*{Crossover temperature}
Class~A dominates Class~B when $2e^{3\beta J}>6e^{-\beta(J+4\gamma w)}$, or equivalently \ $e^{4\beta(J+\gamma w)}>3$. Taking logarithms yields the crossover 
scale directly.

\begin{resultbox}
  \textbf{Result 2 (Crossover scale)}
  \begin{equation}
    \boxed{T^* = \frac{4\,\Jeff}{\ln 3} = \frac{4(J+\gamma w)}{\ln 3}}
    \label{eq:Tstar}
  \end{equation}
  Below $T^*$ aligned configurations dominate the Boltzmann weight, and above it majority-minority configurations prevail. Here, $T^*$ is a finite-system
  crossover, not a genuine phase transition. The latter emerges only in the thermodynamic limit~$N\to\infty$.
\end{resultbox}

Note also that $T^*/T_c = 4/\ln3 \approx 3.64$, hence the crossover of the isolated triad occurs at roughly $3.6\times T_c$, which is well within the disordered 
phase of the network.


\section{Reduction to the Triadic Ising Model}
\label{sec:reduction}

We now demonstrate how systems on triadic hypergraphs map onto the triadic Ising model. We use the full COGENT$^{\textbf{3}}$ system~\cite{salazar2026} as our 
prototypical framework, showing that its critical behavior near the formation transition $T_c^{\mathrm{form}}$ is governed at leading order by the Hamiltonian defined 
in Sec.~\ref{sec:triad}, whose mean-field analysis and composite-operator scaling are developed in Secs.~\ref{sec:mft}--\ref{sec:memory}. All 
COGENT$^{\textbf{3}}$~notation and definitions follow \cite{salazar2026} (specific section references are given inline). 

The argument requires three steps: (i)~symmetry reduction of the formation field; (ii)~collapse of the dynamic topology onto a static all-to-all hypergraph; and 
(iii)~demonstration that the remaining Hamiltonian sectors contribute only irrelevant operators at $T_c^{\mathrm{form}}$.

\subsection*{Symmetry reduction of the formation field}

The formation field $\phi_i(t)\in\mathbb{R}$ in COGENT$^{\textbf{3}}$~carries a global $\mathbb{Z}_2$ symmetry $\phi_i\!\to\!-\phi_i$, inherited from the symmetry of 
the formation correlator $\Pform$ under interchange of formation roles [Sec.~5 therein]. Near $T_c^{\mathrm{form}}$, the effective free energy for $\phi_i$ takes the 
standard $\mathbb{Z}_2$-symmetric Landau form
\begin{equation}
  \mathcal{F}[\phi]
  = \sum_{i}\bigl[a(T)\,\phi_i^{2} + b\,\phi_i^{4}\bigr]
    + \mathcal{F}_{\mathrm{int}}[\phi]
  \label{eq:landau}
\end{equation}
\noindent where $a(T)=a_0(T-T_c^{\mathrm{form}})$ changes sign at $T_c^{\mathrm{form}}$, $b>0$ ensures stability, and $\mathcal{F}_{\mathrm{int}}$ collects the
interaction terms from $H_{\mathrm{group}}$.

The Landau form~\eqref{eq:landau} captures the $\mathbb{Z}_2$-symmetric sector of the formation dynamics. The full formation field $\phi_i(t)$ in COGENT$^{\textbf{3}}$ 
evolves under a stochastic Ginzburg--Landau equation on the induced graph $G^{(2)}(t)$,
\begin{equation}
  \frac{d\phi_i}{dt} = -D_\phi\,(L\phi)_i - \frac{\partial V}{\partial\phi_i} + \sqrt{2\mu\, T_t}\;\xi_i(t)
  \label{eq:GL}
\end{equation}
\noindent with potential
\begin{equation}
  V = \sum_{i} \bigl[a(T)\,\phi_i^{2} + b\,\phi_i^{4}\bigr] + \sum_{\tau\in E^{(3)}} g_\tau \prod_{a\in\tau}\phi_a + \sum_{\{i,j\}\in E^{(2)}} h_{ij}\,(\phi_i\phi_j)^{2}.
  \label{eq:GLpot}
\end{equation}

Note that the cubic term $g_\tau\phi_i\phi_j\phi_k$ (sourced by the triadic topology) is odd under $\phi\to-\phi$ and explicitly breaks the $\mathbb{Z}_2$ symmetry of 
the quartic sector. However, the breaking is subleading near $T_c^{\mathrm{form}}$ because the equilibrium amplitude vanishes as $\phi_{\mathrm{eq}}\to 0$ at the critical 
point, so the cubic contribution to the free energy density scales as $\phi_{\mathrm{eq}}^3\sim|t|^{3/2}$ against $\phi_{\mathrm{eq}}^2\sim|t|$ for the quadratic term. 
The critical behavior is therefore controlled by the $\mathbb{Z}_2$-symmetric quadratic and quartic terms, with the cubic coupling formally irrelevant at the Gaussian 
fixed point.

In the mean-field limit ($N \to \infty$ on the complete hypergraph) critical behavior is determined by the Landau free energy~\eqref{eq:landau}, which depends on the 
$\mathbb{Z}_2$ symmetry and the sign structure of the coefficients $a(T)$ and $b$, not on whether the microscopic field is continuous or discrete. The binary 
discretization $\phi_a \in \{-1,+1\}$ preserves the $\mathbb{Z}_2$ symmetry and produces the same self-consistency equation and Landau expansion near $T_c^{\rm form}$. 
The cubic coupling $g_\tau$ provides corrections of relative order $|t|^{1/2}$ to the leading critical behavior, as established above.

\subsection*{Dynamic topology reduces to all-to-all in the thermodynamic limit}

In COGENT$^{\textbf{3}}$~the active triad set $E^{(3)}(t)$ evolves dynamically. Let $\rho_N$ denote the mean fraction of agent triples forming active triads. We assume 
the extensivity condition $\rho_N \to \rho > 0$ as $N \to \infty$, so that each agent participates in $\sim \rho N^2$ triads. This is the hypergraph analogue of the 
requirement that coordination number diverge with system size in standard mean-field models. If $\rho_N \to 0$, the effective connectivity grows sub-extensively, 
the mean-field self-consistency equation ceases to hold, and no sharp thermodynamic transition occurs. The condition is a regularity assumption on the operating regime, 
and all results in this section hold conditionally on it. 

Under this condition, the effective field on agent $i$ from $H_{\mathrm{group}}$ in the mean-field approximation is
\begin{equation}
  h_i^{\mathrm{eff}}
  = \frac{\Jeff}{N^2}
    \sum_{\tau\ni i}\,\sum_{\substack{j\in\tau \\ j\neq i}}
    \phi_j
  \label{eq:heff}
\end{equation}
\noindent a sum of $n = 2K_i$ spin values, where $K_i = |\{\tau \ni i : \tau \in E^{(3)}\}| \approx \rho N^2/2$ is the number of active triads containing agent $i$.
Writing $\phi_j = m + \delta_j$ with $\mathbb{E}[\delta_j]=0$ and $\delta_j \in [-(1+m),\, 1-m] \subset [-2,2]$, the mean of $h_i^{\mathrm{eff}}$ is $\rho\Jeff m$ and
its fluctuation is $\Delta_i \equiv h_i^{\mathrm{eff}} - \rho\Jeff m = (\Jeff/N^2)\sum_\ell\delta_{j_\ell}$.

Although the formal sum contains $n = 2K_i \approx \rho N^2$ terms, there are only $N-1$ distinct agents. Each $\phi_j$ appears in approximately $c_{ij}\approx\rho N$ 
triads containing $i$. Grouping by distinct agent, define $Y_j = c_{ij}\,\delta_j$ with each $Y_j$ bounded in $[-2\rho N,\, 2\rho N]$. Applying Hoeffding's 
inequality~\cite{hoeffding1963} to the $N-1$ grouped terms\footnote{The grouped $Y_j$ are treated as independent. For weakly correlated spins, concentration persists 
with a modified constant, and a rigorous bound requires a dependent-variable inequality. The leading-order conclusion is however unaffected.} with threshold 
$t = \varepsilon N^2/\Jeff$ gives
\begin{equation}
  P\!\left(|\Delta_i| \geq \varepsilon\right)
  \leq 2\exp\!\left(-\frac{\varepsilon^2\, N}{8\,\rho^2\,\Jeff^2}\right)
  \quad \forall\,\varepsilon > 0.
  \label{eq:hoeffding}
\end{equation}
\noindent For any fixed $\varepsilon > 0$ the right-hand side decays exponentially in $N$. 

The effective field concentrates on $\rho\Jeff m$ as $N\to\infty$, and the dynamic topology is asymptotically equivalent to a complete hypergraph with renormalized 
coupling $\rho\Jeff$. The universality class is unchanged and only $T_c$ is rescaled. Every agent participates in $\binom{N-1}{2} \sim N^2/2$ triads, hence 
$h_i^{\mathrm{eff}} = (\Jeff/N^2)\cdot\rho N^2\cdot m = \Jeff m$ in agreement with~\eqref{eq:HiMF} [see Sec.~\ref{sec:mft}].

Consequently, the dynamic topology enters only through a renormalization of $\Jeff$. The universality class and all critical exponents are independent of $\rho>0$.

\subsection*{Remaining Hamiltonian components are irrelevant at $T_c^{\mathrm{form}}$}

There are three residual components of the COGENT$^{\textbf{3}}$~Hamiltonian that must be accounted for.\footnote{A full renormalization-group analysis of these 
components is deferred to future work.}\\[8pt]
\noindent\textbf{Formation component $H_{\mathrm{form}}$}\\[4pt] 
The GAN parameter variables $(G_\tau^{(m)},D_\tau)$ carry no $\mathbb{Z}_2$ charge under 
$\phi_i\!\to\!-\phi_i$, as they reside in the parameter space of the formation networks and are invariant under sign reversal of the formation field 
[Sec.~4.6.1 therein]. Therefore, they cannot couple to the $\mathbb{Z}_2$ order parameter at odd order. The leading coupling is 
$\phi_{\mathrm{align},\tau}^2\,\phi_i^2$, a $\phi^2$ mass renormalization that shifts $T_c^{\mathrm{form}}$ but does not change the universality class.\\[8pt]
\noindent\textbf{Memory component $H_{\mathrm{mem}}$}\\[4pt] 
The memory field $M_i$ undergoes its own transition at $T_c^{\mathrm{mem}}\neq T_c^{\mathrm{form}}$ 
[Sec.~8.1 therein]. At $T_c^{\mathrm{form}}$ $M_i$ is non-critical, with finite correlation length $\xi_{\mathrm{mem}}<\infty$. Integrating it out generates local 
operators $\delta a\,\phi_i^2+\delta b\,\phi_i^4+\cdots$ that renormalize the Landau coefficients in~\eqref{eq:landau}, without introducing new operators at the 
Ising fixed point.\\[8pt]
\noindent\textbf{Coupling component $H_{\mathrm{coupling}}$}\\[4pt] 
The term $H_{\mathrm{coupling}}=-\sum_\tau\gamma_\tau\phi_{\mathrm{align},\tau}\,m_\tau$ couples the formation 
intensity $\phi_{\mathrm{align},\tau}$ (which is non-critical at $T_c^{\mathrm{form}}$, with gap $\mu_{\mathrm{align}} > 0$) to local opinion coherence $m_\tau$. 
Performing the Gaussian integration over $\phi_{\mathrm{align},\tau}$ at tree level generates an effective pairwise interaction with coupling shift 
$\delta\Jeff = \gamma_\tau^2 / (2\mu_{\mathrm{align}})$, which is finite since $\mu_{\mathrm{align}} > 0$. This shift is absorbed into the renormalized coupling 
$\Jeff$ [Eq.~\eqref{eq:jeff}]. The quantitative accuracy of the pairwise baseline requires $\delta\Jeff / \Jeff \ll 1$.\\[8pt]
Taken together, the three reduction steps above show that the effective theory governing COGENT$^{\textbf{3}}$ near $T_c^{\mathrm{form}}$ is, at leading order, 
the triadic Ising model with coupling $\Jeff$.


\section{Mean-Field Theory}
\label{sec:mft}

Consider now $N$ agents in a complete $3$-uniform hypergraph, with Hamiltonian
\begin{equation}
  \mathcal{H} = -\frac{\Jeff}{N^2}\sum_{\tau\in E^{(3)}} \sum_{\langle ab\rangle\subset\tau}\phi_a\phi_b - h\sum_{i=1}^{N}\phi_i
  \label{eq:Hnet}
\end{equation}
\noindent where the normalization by $N^{-2}$ ensures extensivity. In a mean-field approximation, $\phi_j\to m\equiv\langle\phi_j\rangle$ for all $j\ne i$.
As previously noted, agent~$i$ participates in $\binom{N-1}{2}\sim N^2/2$ triads. Each such triad contributes two terms of the form $\Jeff\phi_i\phi_j/N^2$ to 
$\mathcal{H}$. Replacing $\phi_j\to m$ and summing over all $\sim N^2/2$ triads gives $2\Jeff m\cdot(N^2/2)/N^2=\Jeff m$ in the limit $N\to\infty$. 

The effective single-site Hamiltonian from collecting all contributions is
\begin{equation}
  H_i^{\rm MF} = -(\Jeff m + h)\phi_i + \Order(N^{-1})
  \label{eq:HiMF}
\end{equation}
\noindent from which the self-consistency equation
\begin{equation}
  m = \tanh(\beta\Jeff m + \beta h)
  \label{eq:mf}
\end{equation}
\noindent follows. Linearizing in $m$ and $h$ at $h=0$ reveals the critical condition $\beta_c\Jeff=1$, leading to the next result.
\vspace*{5pt}
\begin{resultbox}
  \textbf{Result 3 (Critical temperature)}
  \begin{equation}
    \boxed{T_c = \Jeff = J+\gamma w}
    \label{eq:Tc}
  \end{equation}
  The gradient coupling strictly raises $T_c$ above the pure-alignment value~$J$, confirming that gradient penalization \emph{promotes} ferromagnetic order.
  Below $T_c$ we have
  \begin{equation}
    m \;\sim\; \sqrt{3}\left(\frac{T_c-T}{T_c}\right)^{1/2} \; \text{as} \quad T\to T_c^-
    \label{eq:m}
  \end{equation}
  with the standard mean-field exponent $\beta_{\rm Ising}^{\rm MF}=1/2$.
\end{resultbox}


\section{Triadic Formation Order Parameter}
\label{sec:order}

The single-spin magnetization $m$ is not the physically natural observable in triadic networks. The emergence of a coherent computational structure within a
triad is measured by the \emph{three-spin correlator}
\begin{equation}
  \Pform = \Bigl\langle\frac{1}{|E^{(3)}|} \sum_{\tau=\{i,j,k\}}\phi_i\phi_j\phi_k\Bigr\rangle_T
  \label{eq:Pform}
\end{equation}
\noindent which is zero in the disordered phase (by $\mathbb{Z}_2$ symmetry) and nonzero whenever a triad achieves coherent configuration. It corresponds to the 
formation order parameter $\Psi_\mathrm{form}$ introduced in~\cite{salazar2026} [Eq.~(95) therein] but derived here from first principles in the Boolean limit
of the group-interaction sector.
\vspace*{-7pt}
\subsection*{Mean-field factorization}
In the $N \to \infty$ limit on a fully connected hypergraph, the standard cumulant expansion for the three-spin correlator is
\begin{equation}
  \langle\phi_i\phi_j\phi_k\rangle
  = m^3 + 3m\underbrace{\langle\phi_i\phi_j\rangle_c}_{\sim\,\Order(N^{-1})}
    + \underbrace{\langle\phi_i\phi_j\phi_k\rangle_c}_{\sim\,\Order(N^{-2})}.
  \label{eq:factorize}
\end{equation}

\begin{resultbox}
  \textbf{Result 4 (Exact factorization)}
  In the thermodynamic limit,
  \begin{equation}
    \Pform = m^3 + \Order(N^{-1})
    \label{eq:cube}
  \end{equation}
  The formation order parameter factorizes exactly to the cube of the single-site magnetization in the thermodynamic limit $N \to \infty$.
\end{resultbox}

\subsection*{Critical exponent}
Substituting $m\sim(T_c-T)^{1/2}$ from \eqref{eq:m} into \eqref{eq:cube} yields:

\begin{resultbox}
  \textbf{Result 5 (TF critical exponent)}
  \begin{equation}
    \boxed{\Pform \;\sim\; (T_c-T)^{3/2} \;\; \text{as} \;\; T\to T_c^-}
    \label{eq:Pscale}
  \end{equation}
  defining the \textbf{\emph{Triadic Formation (TF) critical exponent}}
  \begin{equation}
    \bTF = 3\,\beta_{\rm Ising}^{\rm MF} = \tfrac{3}{2}.
    \label{eq:beta}
  \end{equation}
\end{resultbox}

This exponent does not coincide with the primary order-parameter exponent of mean-field Ising ($\beta_\mathrm{Ising}^\mathrm{MF} = \tfrac{1}{2}$), mean-field XY 
($\beta_\mathrm{XY}^\mathrm{MF} = \tfrac{1}{2}$), or mean-field Heisenberg ($\beta_\mathrm{Heis}^\mathrm{MF} = \tfrac{1}{2}$) universality classes, all of which share 
$\beta^\mathrm{MF} = 1/2$ regardless of the spin symmetry group (since mean-field theory is insensitive to the number of order-parameter components). However, it is 
precisely the composite-operator exponent $n\beta_\mathrm{Ising}^\mathrm{MF}$ evaluated at $n = 3$ which follows from 
$\Delta_{\Psi_\mathrm{form}} = 3\Delta_\phi$~\cite{goldenfeld2018, stanley1971}.

Equation~\eqref{eq:beta} is an exact consequence of the three-body structure. Coherent formation requires three spins to align simultaneously, imposing a cubic 
relation between $\Psi_\mathrm{form}$ and the underlying magnetization. The exponent $\beta_\mathrm{TF} = 3/2$ is an algebraic consequence of the composite-operator
structure, and any order parameter constructed as the cube of a mean-field Ising mode inherits this effective exponent via
$\Delta_{\Psi_\mathrm{form}} = 3\Delta_\phi$.

\subsection*{Absence of susceptibility divergence}
Let $h_3$ couple extensively to the formation order parameter as $\mathcal{H}\to\mathcal{H} - h_3 N\Pform$. The thermodynamically consistent, intensive linear response 
is
\begin{equation}
  \chiTF \equiv \frac{\partial\Pform}{\partial h_3}\bigg|_{h_3=0}
  = \beta\,N\,\mathrm{Var}(\Pform).
  \label{eq:chiTF}
\end{equation}

It parallels the standard Ising definition $\chi_\mathrm{Ising} = \beta N\,\mathrm{Var}(m)$, with $\Pform$ playing the role of order parameter. Both are
$\Order(1)$ in $N$: since $\mathrm{Var}(m) = \chi_\mathrm{Ising}/(\beta N) = \Order(N^{-1})$, the factor of $N$ in~\eqref{eq:chiTF} cancels exactly, thus
$\chiTF = \beta N\cdot(3m^2)^2\mathrm{Var}(m) = 9m^4\chi_\mathrm{Ising}$, which is genuinely intensive.

Using $\Pform=m^3$ and the delta-method approximation $\mathrm{Var}(m^3)\approx(3m^2)^2\mathrm{Var}(m)$, together with $m^2\sim(T_c-T)$ and
$\chi_\mathrm{Ising}\sim(T_c-T)^{-1}$ near $T_c$.

\begin{resultbox}
  \textbf{Result 6 (No susceptibility divergence)}
  \begin{equation}
  \begin{split}
    \chiTF
    = 9m^4\chi_\mathrm{Ising}
    &\sim (T_c-T)^{2}\cdot(T_c-T)^{-1} \\
    &= (T_c-T)^{+1} \to 0 \;\; \text{as} \;\; T\to T_c^-
  \end{split}
  \label{eq:chi}
  \end{equation}
  The formation susceptibility \emph{vanishes} at $T_c$, giving $\gTF = -1$ as the effective scaling exponent. For $T>T_c$, the $h_3$-induced correction to the 
  effective single-site field is $3h_3 m^2$ (from differentiating the coupling $-h_3 N\Pform = -h_3 N m^3$ with respect to $\phi_i$ in the mean-field approximation), 
  which vanishes when $m=0$. As a result, the mean-field linear response of $\Pform = m^3$ to $h_3$ is zero in the disordered phase, and $\chiTF=0$ from above. There 
  is no divergence at $T_c$ in response to the formation field.
\end{resultbox}

The delta-method captures the leading power law. Near $T_c$ fluctuations in $m$ modify the prefactor but not the exponent $\gamma_{\mathrm{TF}}=-1$.

As shown in the next section, the scaling exponent $\gTF = -1$ is reproduced by the field-theoretic two-point function. This is in sharp contrast to Ising criticality 
($\chi_{\rm Ising}\sim |T-T_c|^{-1}$ diverges on both sides). In principle, the non-diverging, monotone vanishing of $\chiTF$ at $T_c$ (which approaches $T_c$ with a 
kink rather than a divergence) constitutes a signature of the TF scaling regime (see Sec.~\ref{sec:disc}).

\subsection*{Scaling consistency}
The exponents $\beta_{\rm TF}$ and $\gamma_{\rm TF}$ are defined via $\Psi_{\rm form} = m^3$. The specific heat exponent 
$\alpha_{\rm TF} \equiv \alpha_{\rm Ising}^{\rm MF} = 0$ since $\alpha$ characterizes the free-energy singularity $f_{\rm sing} \sim |t|^{2-\alpha}$, which is a 
property of the partition function and independent of the choice of order-parameter observable. The Rushbrooke equality
\begin{equation}
    \alpha_\mathrm{TF} + 2\beta_\mathrm{TF} + \gamma_\mathrm{TF}
    = 0 + 3 + (-1) = 2 \label{eq:rushbrooke}
\end{equation}
\noindent is satisfied by algebraic construction and constitutes a self-consistency check on the exponent definitions, not an independent thermodynamic test.

A complementary check is provided by the field-theoretic two-point function at criticality. At the upper critical dimension $d_c = 4$, the scalar field 
$\phi$ has mean-field scaling dimension $\Delta_\phi = 1$. The composite operator $\Psi_\mathrm{form} = \phi^3$ then has dimension $\Delta_{\Psi_\mathrm{form}} = 3$. 
Its momentum-space susceptibility scales as
\begin{equation}
    \chi_\mathrm{TF}(q) \sim |q|^{2\Delta_{\Psi_\mathrm{form}} - d}
    = |q|^{6-4} = |q|^2 
    \label{eq:chiTF_q}
\end{equation}
\noindent so the uniform susceptibility $\chi_\mathrm{TF}(q{=}0) \sim \xi^{d - 2\Delta_{\Psi_\mathrm{form}}} = \xi^{-2} \sim t^{+1}$, giving $\gamma_\mathrm{TF} = -1$. 
This is the momentum-space restatement of the mean-field computation, not an independent derivation: below $d_c$ the local operator $\phi^3$ mixes with $\phi$, and the 
long-distance behavior of the \emph{separated} correlator $\Pform$ is instead governed by cluster decomposition (Sec.~\ref{sec:beyondMF})~\cite{zinn_justin2002}.


\section{Three-Body Coupling and the Formation Phase Diagram}
\label{sec:threebody}

The results of Secs.~\ref{sec:mft}--\ref{sec:order} are derived in the $\lambda = 0$ limit, where the three-body coupling in $H_{\mathrm{group}}$ is absent and the 
Hamiltonian is exactly pairwise. We now restore the three-body term and determine its effect on the formation transition.

\subsection*{Mean-field theory with three-body coupling}

On the complete $3$-uniform hypergraph the Hamiltonian including the three-body energy is
\begin{equation}
  \mathcal{H}_\lambda = -\frac{\Jeff}{N^2}\sum_{\tau\in E^{(3)}}
      \sum_{\langle ab\rangle\subset\tau}\phi_a\phi_b
    \;-\; \frac{\lambda}{N^2}\sum_{\tau\in E^{(3)}}
      \prod_{a\in\tau}\phi_a
    \;-\; h\sum_{i=1}^{N}\phi_i
  \label{eq:Hlambda}
\end{equation}
\noindent where the $N^{-2}$ normalization ensures extensivity for both terms.In the mean-field approximation, the three-body coupling contributes an additional term 
$\tfrac{1}{2}\lambda m^2$ to the effective field on agent~$i$ (each of the $\sim N^2/2$ triads containing $i$ contributes $\lambda m^2/N^2$ after replacing 
$\phi_j,\phi_k\to m$). The effective single-site Hamiltonian generalizes~\eqref{eq:HiMF} to
\begin{equation}
  H_i^{\mathrm{MF}} 
  = -\bigl(\Jeff\, m + \tfrac{\lambda}{2}\, m^2 + h\bigr)\phi_i 
    + \Order(N^{-1})
  \label{eq:HiMFlambda}
\end{equation}
\noindent and the self-consistency equation becomes
\begin{equation}
  m = \tanh\!\Bigl[\beta\Bigl(\Jeff\, m + \tfrac{\lambda}{2}\, m^2 + h\Bigr)\Bigr].
  \label{eq:mf_lambda}
\end{equation}

Under $m\to -m$ at $h=0$, the argument of the hyperbolic tangent maps to $\beta(-\Jeff m + \tfrac{\lambda}{2}m^2)$, which is not equal to 
$-\beta(\Jeff m + \tfrac{\lambda}{2}m^2)$. The three-body coupling breaks the $\mathbb{Z}_2$ symmetry of the self-consistency equation for any $\lambda\neq 0$.

\subsection*{Landau free energy and first-order transition}

The mean-field free energy per spin is
\begin{equation}
  f(m) = -\frac{\Jeff}{2}\,m^2 - \frac{\lambda}{6}\,m^3 + T\biggl[\frac{1{+}m}{2}\ln\frac{1{+}m}{2} + \frac{1{-}m}{2}\ln\frac{1{-}m}{2}\biggr] - hm.
  \label{eq:f_lambda}
\end{equation}
Expanding the entropy for small $m$ gives the Landau form
\begin{equation}
  f(m) = f_0 + \frac{T - \Jeff}{2}\,m^2 - \frac{\lambda}{6}\,m^3 + \frac{T}{12}\,m^4 + \Order(m^6).
  \label{eq:landau_lambda}
\end{equation}

The cubic coefficient $\lambda/6$ is nonzero for any $\lambda\neq 0$, and the free energy is asymmetric as $f(m)\neq f(-m)$. This is the standard Landau signature of a 
first-order transition~\cite{goldenfeld2018}.

At $h = 0$, the equilibrium condition $\partial f/\partial m = 0$ yields either $m = 0$ (disordered phase) or
\begin{equation}
  (T - \Jeff) - \frac{\lambda}{2}\,m + \frac{T}{3}\,m^2 = 0.
  \label{eq:equil_lambda}
\end{equation}

The first-order transition occurs where the ordered and disordered minima have equal free energy, $f(m^*) = f(0)$. Imposing this condition simultaneously 
with~\eqref{eq:equil_lambda} gives the magnetization at the transition
\begin{equation}
  m^* = \frac{\lambda}{T_t}
  \label{eq:mstar}
\end{equation}
\noindent and the transition temperature
\begin{equation}
  T_t^2 - \Jeff\, T_t - \frac{\lambda^2}{6} = 0
  \label{eq:Tt_quad}
\end{equation}
\noindent the positive root of which is the following result.

\begin{resultbox}
  \textbf{Result 7 (First-order transition with three-body coupling)}
  For $\lambda\neq 0$, the formation transition is first-order, occurring at
  \begin{equation}
    \boxed{T_t = \frac{\Jeff + \sqrt{\Jeff^2 + \tfrac{2}{3}\lambda^2}}{2}}
    \label{eq:Tt}
  \end{equation}
  with magnetization discontinuity $\Delta m = \lambda/T_t$ and latent heat per spin $L = \lambda^2/(2T_t)$. For weak three-body 
  coupling ($\lambda/\Jeff\ll 1$)
  \begin{equation}
    T_t \approx \Jeff + \frac{\lambda^2}{6\,\Jeff} + \Order(\lambda^4) \quad \text{and} \quad \Delta m \approx \frac{\lambda}{\Jeff}.
    \label{eq:Tt_weak}
  \end{equation}
  The transition occurs above the pairwise critical temperature $T_c = \Jeff$. The continuous Ising critical point is recovered only in the limit $\lambda\to 0$.
\end{resultbox}

\subsection*{Phase diagram and critical point}

\begin{figure*}[t]
    \centering
    \includegraphics[width=0.6\textwidth]{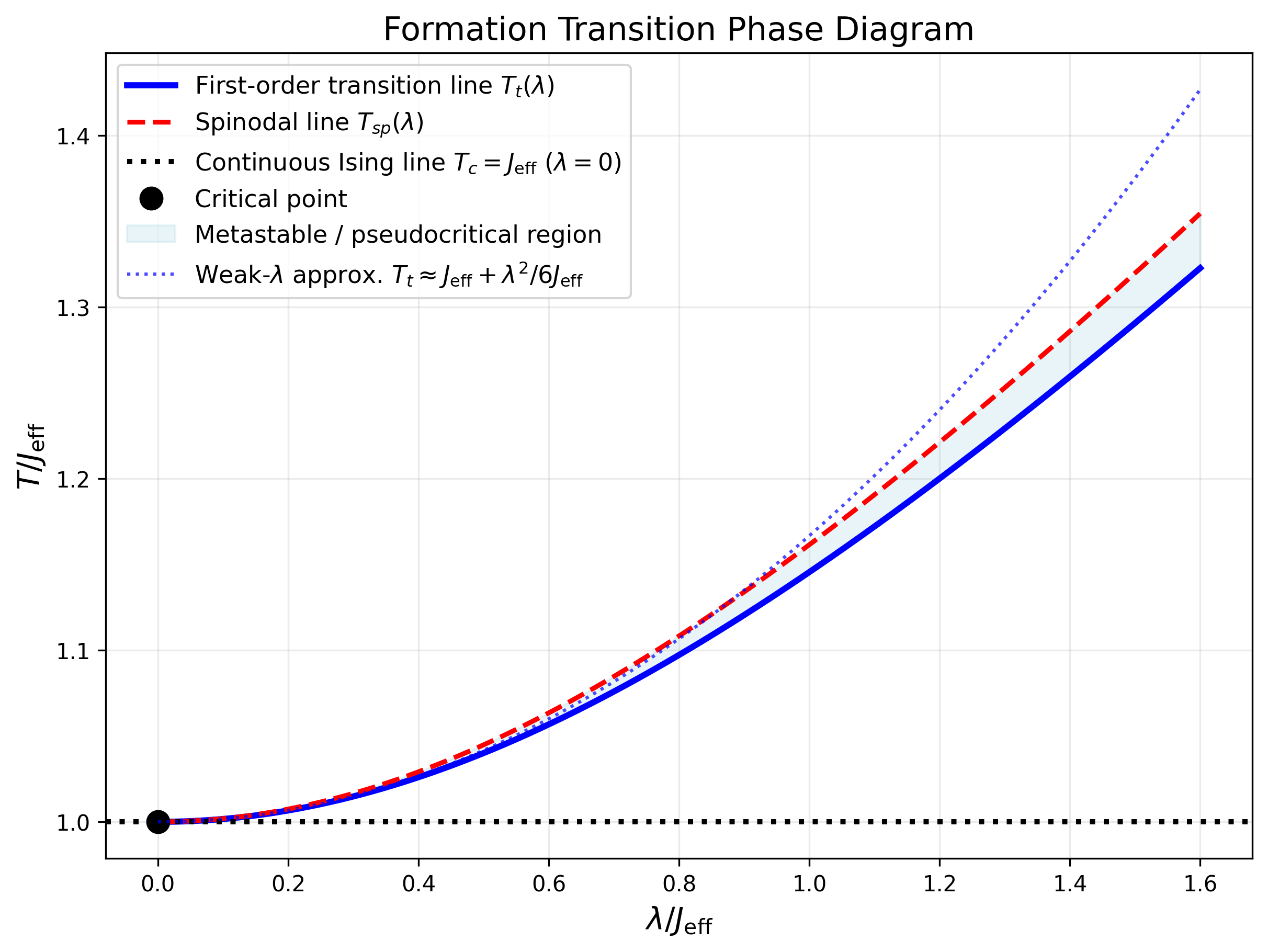}
    \captionsetup{width=0.7\textwidth, justification=justified}
    \caption{Formation transition in the $(\lambda/J_{\rm eff}, T/J_{\rm eff})$ plane. The black dotted line marks the continuous 
    transition $T_c = J_{\rm eff}$ at $\lambda=0$, which ends at the critical endpoint (black circle). For $\lambda>0$ the transition becomes first-order (solid blue 
    line $T_t(\lambda)$, Result 7). The red dashed line is the spinodal $T_{sp}(\lambda)$. Light blue shading shows the metastable/pseudocritical region. The thin 
    blue dotted line shows the weak-coupling approximation $T_t \approx J_{\rm eff} + \lambda^2/(6J_{\rm eff})$.}
    \label{fig:phase_diagram}
\end{figure*}

The structure of the phase diagram in the $(\lambda/\Jeff,\; T/\Jeff)$ plane (see Figure~\ref{fig:phase_diagram} above) is as follows.
\begin{itemize}
  \item At $\lambda = 0$ the transition is continuous (second order) at $T_c = \Jeff$, with Ising mean-field exponents. The magnetization vanishes continuously as 
  $m\sim(T_c - T)^{1/2}$, and the composite-operator scaling of Sec.~\ref{sec:order} applies: $\bTF = 3/2$ and $\gTF = -1$.
  \item For any $\lambda > 0$ the transition is first-order at $T_t(\lambda) > \Jeff$, with a finite discontinuity $\Delta m = \lambda/T_t$. The continuous scaling 
  $m\sim (T_c - T)^{1/2}$ does not hold at the transition itself.
  \item The line of first-order transitions $T_t(\lambda)$ terminates at the point $(\lambda, T) = (0,\, \Jeff)$, which is the \emph{critical endpoint} of the phase 
  diagram. There the first-order discontinuity vanishes ($\Delta m\to 0$), the latent heat vanishes ($L\to 0$), and the transition recovers the continuous 
  $\mathbb{Z}_2$ Ising critical point.
\end{itemize}

\subsection*{Pseudocritical regime}

For $\lambda/\Jeff \ll 1$ the first-order transition is weak: $\Delta m \sim \lambda/\Jeff$ and $L\sim\lambda^2/\Jeff$. In this regime the system exhibits 
\emph{pseudocritical} behavior over a range of temperatures. The spinodal (where the metastable ordered minimum disappears) occurs at temperature
\begin{equation}
  T_{\mathrm{sp}}^2 - \Jeff\, T_{\mathrm{sp}} - \frac{3\lambda^2}{16} = 0,
  \label{eq:Tsp}
\end{equation}
\noindent giving $T_{\mathrm{sp}} \approx \Jeff + 3\lambda^2/(16\Jeff)$ for small $\lambda$. The width of the metastable region is
\begin{equation}
  T_{\mathrm{sp}} - T_t \approx \frac{\lambda^2}{48\,\Jeff}
  \label{eq:metastable_width}
\end{equation}
\noindent which is quadratically small in $\lambda/\Jeff$. Outside this narrow window, the continuous Ising scaling $m\sim |t|^{1/2}$ is a controlled approximation, 
with deviations confined to a temperature interval of width $\sim \lambda^2/\Jeff$ around $T_t$. 

Within this pseudocritical regime, the composite-operator results $\bTF = 3/2$ and $\gTF = -1$ remain valid to leading order. The pairwise baseline of 
Secs.~\ref{sec:mft}--\ref{sec:order} is therefore the $\lambda = 0$ critical limit, and its predictions are quantitatively accurate whenever $\lambda/\Jeff$ is 
sufficiently small that the pseudocritical window is broad compared to the first-order discontinuity.


\section{Composite-Operator Scaling Beyond Mean Field}
\label{sec:beyondMF}

The mean-field results of Sec.~\ref{sec:order} are derived at the upper critical dimension $d_c = 4$. We now show that the composite-operator exponent \emph{relations} 
hold exactly in arbitrary dimension $d$, and determine the dimension at which the vanishing-susceptibility signature changes character.

\subsection*{Exact factorization from cluster decomposition}

The formation correlator~\eqref{eq:Pform} is a spatial average of the multi-site product $\phi_i\phi_j\phi_k$ over distinct agents $i\neq j\neq k$. For any pair of 
separated sites in a translation-invariant system, the connected two-point function decays as $\langle\phi_i\phi_j\rangle_c \sim |r_{ij}|^{-(d-2+\eta)}$. On a 
fully connected graph, $\langle\phi_i\phi_j\rangle_c = \Order(N^{-1})$. In either case, the cumulant expansion
\begin{equation}
  \langle\phi_i\phi_j\phi_k\rangle = m^3 + 3m\,\langle\phi_i\phi_j\rangle_c + \langle\phi_i\phi_j\phi_k\rangle_c
  \label{eq:cluster}
\end{equation}
\noindent has connected contributions that are sub-extensive when summed over all $\Order(N^3)$ triads and normalized by $|E^{(3)}|$. The factorization
\begin{equation}
  \Pform = m^3 + \text{sub-extensive corrections}
  \label{eq:exact_factor}
\end{equation}
\noindent is therefore exact in the thermodynamic limit, in \emph{any} dimension $d$, provided $E^{(3)}$ is the complete set of triples or a dense set with 
$|E^{(3)}|=\Theta(N^3)$ for which almost every triad has mutually divergent pairwise separations. Equivalently, the connected pairwise correlator is 
subextensive when summed over $E^{(3)}$. It is then a consequence of cluster decomposition under this density condition, not of mean-field theory. The multi-site 
structure alone does not suffice.

In the dense regime, the formation correlator $\Pform$ is not a local operator: it is an average over products of fields at \emph{separated} sites. The operator mixing 
that generates the anomalous dimension of $\phi^3(x)$ is a short-distance effect that does not contribute to this multi-site average in the thermodynamic limit.

For an extensive set of bounded-diameter (local) triads, $|E^{(3)}|=\Theta(N)$, the connected pairwise correlator $\langle\phi_i\phi_j\rangle_c$ at fixed short 
separation stays bounded away from zero as $T\to T_c^-$ (it does not vanish with $m$). The cross term $m\langle\phi_i\phi_j\rangle_c\sim|t|^{\beta}$ then dominates 
$m^3\sim|t|^{3\beta}$, giving $\bTF=\beta_{\mathrm{Ising}}$. The relation $\bTF=3\beta_{\mathrm{Ising}}$ is therefore specific to the non-local regime defined above.

\subsection*{Exact exponent relations}

Since $\Pform = m^3$ holds exactly, the TF exponents are determined algebraically by the Ising exponents in any dimension, that is
\begin{equation}
  \bTF = 3\,\beta_{\mathrm{Ising}} \quad \text{with} \quad \gTF = \gamma_{\mathrm{Ising}} - 4\,\beta_{\mathrm{Ising}}.
  \label{eq:exact_relations}
\end{equation}
\noindent The susceptibility relation follows from $\chiTF = 9m^4 \chi_{\mathrm{Ising}}$ (see Result~6) with $m^4\sim |t|^{4\beta}$ and 
$\chi_{\mathrm{Ising}}\sim |t|^{-\gamma}$, giving $\chiTF\sim |t|^{4\beta - \gamma}$ and hence the second relation. The Rushbrooke equality is automatically preserved, 
as
\begin{equation}
  \alpha_{\mathrm{TF}} + 2\bTF + \gTF 
  = \alpha + 6\beta + (\gamma - 4\beta) 
  = \alpha + 2\beta + \gamma = 2.
  \label{eq:rushbrooke_exact}
\end{equation}

\begin{resultbox}
  \textbf{Result 8 (Exact composite-operator exponent relations)}
  The relations
  \begin{equation}   
    \bTF = 3\beta_{\mathrm{Ising}} \quad \text{and} \quad \gTF = \gamma_{\mathrm{Ising}} - 4\beta_{\mathrm{Ising}}
  \end{equation}
  \noindent hold exactly in arbitrary dimension for any system in the Ising universality class, on graphs for which the connected pairwise correlator is subextensive 
  when summed over $E^{(3)}$ (see above). They follow from the cluster decomposition of the multi-site correlator and do not require mean-field theory. These are 
  rigidity relations, as the TF exponents are not independent but determined by the Ising exponents of the underlying order parameter.
\end{resultbox}

\subsection*{One-loop $\epsilon$-expansion}

Using the standard one-loop Wilson--Fisher results for the Ising model in $d = 4 - \epsilon$ dimensions~\cite{goldenfeld2018, zinn_justin2002}
\begin{equation}
  \beta_{\mathrm{Ising}} = \tfrac{1}{2} - \tfrac{\epsilon}{6} + \Order(\epsilon^2) \quad \text{with} \quad
  \gamma_{\mathrm{Ising}} = 1 + \tfrac{\epsilon}{6} + \Order(\epsilon^2)
  \label{eq:ising_eps}
\end{equation}
\noindent the composite-operator exponents become $\bTF = \tfrac{3}{2} - \tfrac{\epsilon}{2} + \Order(\epsilon^2)$ with 
$\gTF = -1 + \tfrac{5\epsilon}{6} + \Order(\epsilon^2)$. At $\epsilon = 0$ ($d = 4$) the mean-field values $\bTF = 3/2$ and $\gTF = -1$ are recovered. 
At $\epsilon = 1$ ($d = 3$) the one-loop estimates give $\bTF \approx 1$ and $\gTF \approx -1/6$.

Using the numerical Ising exponents in $d = 3$ ($\beta \approx 0.3265$, $\gamma \approx 1.2372$) the exact relations~\eqref{eq:exact_relations} yield~\footnote{The 
exponents are derived from the high-precision conformal bootstrap results $\Delta_\sigma = 0.518154(15)$ and $\Delta_\epsilon = 1.41267(13)$ in \cite{ElShowk2014}, 
via the scaling relations $\beta = \frac{\nu}{2}(d-2+\eta)$ and $\gamma = (2-\eta)\nu$ with $\eta = 2(\Delta_\sigma - 1/2)$ and $\nu = 1/(3 - \Delta_\epsilon)$.}
\begin{equation}
  \bTF^{(d=3)} = 3(0.3265) \approx 0.980 \quad \text{and} \quad \gTF^{(d=3)} = 1.2372 - 4(0.3265) \approx -0.069.
  \label{eq:TF_3d}
\end{equation}
\noindent The formation susceptibility still vanishes at $T_c$ in three dimensions, but weakly $\chiTF\sim |t|^{0.069}$ compared to $|t|^{1}$ in mean field.

\subsection*{Critical dimension for the vanishing susceptibility}

The vanishing-susceptibility signature ($\gTF < 0$) persists as long as $\gamma_{\mathrm{Ising}} < 4\beta_{\mathrm{Ising}}$. Using hyperscaling ($\gamma = (2-\eta)\nu$, 
$\beta = \nu(d-2+\eta)/2$), the condition $\gTF = 0$ gives
\begin{equation}
  d^* = 3 - \tfrac{3}{2}\eta(d^*).
  \label{eq:dstar}
\end{equation}

Since $\eta > 0$ for all $d < 4$, the critical dimension satisfies $d^* < 3$. The vanishing susceptibility therefore holds in all dimensions $d \geq 3$, including the 
physically relevant case $d = 3$. This dichotomy is subject to the same dense/non-local condition as Result~8. For local triads, $\bTF=\beta$ and it does not apply.

In $d = 2$ the exact Onsager exponents ($\beta = 1/8$, $\gamma = 7/4$) give~\footnote{Refer to \cite{stanley1971, onsager1944}.}
\begin{equation}
  \gTF^{(d=2)} = \tfrac{7}{4} - 4\cdot\tfrac{1}{8} = \tfrac{5}{4} > 0
  \label{eq:TF_2d}
\end{equation}
\noindent hence the formation susceptibility \emph{diverges} in two dimensions. 

The vanishing-susceptibility signature is therefore not a universal feature of the TF composite-operator regime: it holds above a critical dimension $d^*\lesssim 3$ and 
fails below it. The transition from diverging to vanishing formation susceptibility as $d$ increases through $d^*$ constitutes a qualitative change in the fluctuation 
structure of the triadic correlator.


\section{Memory-Modified Dynamical Exponent}
\label{sec:memory}

The architecture of COGENT$^{\textbf{3}}$~assigns each agent a memory field. To capture the history-dependent dynamics without resolving the full microscopic 
fast-variable dynamics, we describe this field using a phenomenological Mori--Zwanzig memory ansatz $K(t)$~\cite{mori1965, zwanzig1960, zwanzig2001}. 
Following~\cite{salazar2026}, in the vicinity of a critical point this memory kernel modifies the dynamical exponent $z$ (relating relaxation time to correlation 
length, $\tau_{\rm relax}\sim\xi^z$) to
\begin{equation}
  z_\alpha = z_\alpha^{(0)} + \gamma_K\int_0^\infty K(t)\,t^{\theta_\alpha-1}\,\mathrm{d}t
  \label{eq:zdyn}
\end{equation}
\noindent with $\theta_\alpha$ the autocorrelation decay exponent of the observable $\alpha$. For $\Psi_\mathrm{form}=m^3$, the exponent $\theta_\mathrm{TF}$ is 
obtained by analyzing the six-point correlation function in the disordered phase $T>T_c$, where Gaussian fluctuations dominate.

The calculation is straightforward. In the disordered phase $T > T_c$, the free energy is dominated by the quadratic term $F \sim \tfrac{1}{2}N a(T) m^2$ with 
$a(T) = a_0(T - T_c) > 0$. Hence, the $m$-fluctuations are Gaussian to leading order in $N^{-1}$, with $\langle m^2\rangle = [Na(T)]^{-1}$, and Isserlis' theorem 
applies exactly.

Denoting $C(t) \equiv \langle m(t)m(0)\rangle_c$, the six-point function factorizes as
\begin{equation}
    \langle m(t)^3 m(0)^3 \rangle
    = 6\,C(t)^3 + 9\,\langle m^2\rangle^2\,C(t).
    \label{eq:wick}
\end{equation}

Both terms scale identically in $N$ because $\langle m^2\rangle^2 \sim N^{-2}$ and $C(t) \sim \langle m^2\rangle g(t) \sim N^{-1}g(t)$, giving
$6C(t)^3 \sim N^{-3} g(t)^3$ and $9\langle m^2\rangle^2 C(t) \sim N^{-3} g(t)$. 

The linear term in $C(t)$ is not suppressed by any additional power of $N$, as it is of the same order as the cubic term. For a single-exponential correlation 
$C(t)=\langle m^2\rangle e^{-t/\tau_0}$, the autocorrelation of $\Psi_\mathrm{form}$ becomes
\begin{equation}
    \langle \Psi_\mathrm{form}(t)\Psi_\mathrm{form}(0)\rangle
    \propto \langle m^2\rangle^3\bigl[6e^{-3t/\tau_0} + 9e^{-t/\tau_0}\bigr].
    \label{eq:corr}
\end{equation}

At long times, $e^{-t/\tau_0}$ dominates $e^{-3t/\tau_0}$ so the $\Pform$ autocorrelation is controlled by the same exponential scale as $m$. This
off-critical argument (valid for all $T>T_c$) shows that $\Pform$ and $m$ share the same long-time decay mode. By continuity as $T\to T_c^+$, where
the exponential relaxation time $\tau_0\sim\xi^z$ diverges and the decay crosses over to a critical power law $C(t)\sim t^{-\theta_0}$, the composite correlator 
inherits the same power-law exponent. As a result, $\theta_\mathrm{TF} = \theta_0$, the critical autocorrelation exponent of the single-spin magnetization $m$.

For the canonical single-exponential kernel $K(t) = \tau_K^{-1}e^{-t/\tau_K}$, the integral in~\eqref{eq:zdyn} evaluates to $\tau_K^{\theta_0-1}\Gamma(\theta_0)$ 
via the Mellin transform.

\begin{resultbox}
  \textbf{Result 9 (Memory correction to dynamical exponent)}
  \begin{equation}
    \boxed{\zTF = z_0^{(0)} + \gamma_K\,\tau_K^{\theta_0-1}\,\Gamma(\theta_0)}
    \label{eq:z}
  \end{equation}
  The memory correction $\Delta z = \gamma_K\tau_K^{\theta_0-1}\Gamma(\theta_0)$ is (i)~monotone \emph{increasing} in $\tau_K$ when $\theta_0>1$ and monotone
  \emph{decreasing} when $0<\theta_0<1$, so the direction of memory-induced slowing depends on the bare autocorrelation exponent; (ii)~analytically tunable,
  making $\zTF$ a continuously adjustable parameter; and (iii)~of sign fixed by $\gamma_K$: for $\gamma_K>0$ (dissipative memory) $\Delta z>0$ and memory
  enhances critical slowing, while for $\gamma_K<0$ (anti-persistent memory) $\Delta z<0$ and memory accelerates dynamics.
\end{resultbox}

For $\theta_0>1$, in the limit $\tau_K\to0$ (memoryless), $\Delta z\to0$, and the standard exponent is recovered. When $\tau_K\to\infty$, $\Delta z$ grows without 
bound \emph{within the ansatz}. This limit lies outside the small-correction regime in which Eq.~\eqref{eq:z} is controlled, and we do not interpret it as a genuine 
arrest transition. For $0<\theta_0<1$ the limits are reversed (long memory suppresses the dynamical correction while short memory amplifies it). 

The bare exponent $z_0^{(0)}$ and the critical autocorrelation exponent $\theta_0$ are not free parameters, as they are both determined by the dynamical universality 
class and the spatial dimension~\cite{zinn_justin2002}. For non-conserved relaxational dynamics $z_0^{(0)} = 2$ at mean field, so if a conservation law constrains the 
critical mode, $z_0^{(0)} = 4$. The appropriate value for the formation sector depends on whether the formation norm conservation law [Sec.~3 of~\cite{salazar2026}] 
projects onto the critical mode, a question deferred to future work. The only genuinely model-dependent quantities in Eq.~\eqref{eq:z} are the memory coupling 
$\gamma_K$ and the memory timescale $\tau_K$, which characterize the memory kernel and must be measured or specified for a given system.

\newpage
\section{Discussion}
\label{sec:disc}

The nine core results presented above characterize the properties of the TF composite scaling regime. A comparison against their classical counterparts is provided in 
Table~\ref{tab:exponents}. In addition, our theoretical mean-field derivations are empirically validated via Monte Carlo simulations on random 3-uniform hypergraphs, 
with methodology and numerical results detailed in the Appendix.

\begin{table}[h]
  \centering
  \caption{Critical exponents of the TF scaling regime (this work) versus standard mean-field classes. Exponents for TF refer to the
  composite operator $\Pform=m^3$. Rushbrooke is satisfied in all cases [refer to equation~\eqref{eq:rushbrooke}].}
  \label{tab:exponents}
  \medskip
  \begin{tabular*}{0.6\linewidth}{@{\extracolsep{\fill}}lccc@{}}
    \toprule
    \multirow{2}{*}{Exponent} & TF          & MF       & MF              \\
                              & (this work) & Ising    & XY | Heisenberg \\
    \midrule
    $\beta$                   & $3/2$       & $1/2$    & $1/2$           \\
    $\gamma$                  & $-1$        & $+1$     & $+1$            \\
    $\alpha$                  & $0$         & $0$      & $0$             \\
    $\nu$                     & $1/2$       & $1/2$    & $1/2$           \\
    $\alpha+2\beta+\gamma$    & $2$         & $2$      & $2$             \\
    \bottomrule
  \end{tabular*}
\end{table}

\noindent Four remarks are now in order, outlined below.\\[8pt]
\noindent\textbf{Exactness of $\bTF=3/2$}\\[4pt]
The result is not perturbative but follows from the exact factorization $\Pform=m^3$ in the $N\to\infty$ mean-field limit,
where connected three-body correlations are suppressed by $\Order(N^{-2})$ (Result~4). Corrections at finite $N$ modify the exponent by terms $\sim N^{-1}$ arising 
from the connected three-body correlator.\\[8pt]
\noindent\textbf{Negative effective $\gTF$}\\[4pt] 
The susceptibility $\chiTF\to0$ at $T_c$ is an exact consequence of the cubic relation~\eqref{eq:cube} and is physically
interpretable. Three agents must be coherently perturbed by the field $h_3$ to produce a response, and the probability of joint triadic alignment vanishes as
the system approaches the disordered phase from below. This feature is in principle separable from Ising criticality: whereas $\chi_\mathrm{Ising}$ diverges symmetrically 
on both sides of $T_c$, a direct measurement of $\chi_\mathrm{TF}$ requires applying a field $h_3$ conjugate to the three-body product $\phi_i\phi_j\phi_k$ and recording 
the response of $\Psi_\mathrm{form}$. In a system with triadic interactions, $h_3$ corresponds to a coordinated triadic bias. Its experimental identification and the 
resulting non-diverging susceptibility profile provide a concrete protocol for distinguishing TF from scalar-Ising criticality, deferred to future work.\\[8pt]
\noindent \textbf{Tunable dynamical scaling}\\[4pt] 
Result~9 shows that the TF regime exhibits a continuously tunable dynamical exponent parametrized by $\tau_K$, 
qualitatively analogous (although mechanistically distinct) to the Kosterlitz--Thouless line in two-dimensional systems.\\[8pt]
\noindent \textbf{The validity of mean-field theory and upper critical dimension}\\[4pt] 
The mean-field results (Results~3--9) are exact in the $N\to\infty$ limit of the 
fully connected hypergraph. The composite-operator exponent relations extend beyond mean field as exact consequences of cluster decomposition (Result~8), with 
the vanishing-susceptibility signature persisting in $d\geq 3$ but failing in $d = 2$ (Sec.~\ref{sec:beyondMF}).\\[8pt]
The present work characterizes the TF scaling regime at the mean-field level. Natural extensions include the following:
\begin{itemize}
  \item Fluctuation corrections in spatial dimension $d<d_c=4$ via $\epsilon$-expansion renormalization-group methods, verifying that the Ginzburg criterion 
  argument of Sec.~\ref{sec:disc} extends to the composite operator $\phi^3$ at the Wilson-Fisher fixed point, and determining whether the composite-operator relation 
  $\beta_\mathrm{TF}^\mathrm{RG}=3\beta_\mathrm{Ising}^\mathrm{RG}$ is preserved below $d_c$;
  \item Extension to the COGENT$^{\textbf{3}}$~role-transition sector, which introduces a $\mathbb{Z}_3$ (Potts-like) component coupled to the Ising $\Pform$, 
  potentially giving rise to a multicritical point; and 
  \item Identification of measurable experimental signatures in dynamic triadic hypergraphs, where the non-diverging susceptibility and the $\beta_{\rm TF}=3/2$ scaling 
  provide direct tests of the TF scaling regime.
\end{itemize}


\section{Conclusions}
\label{sec:conc}

In this paper, we have derived two analytically exact results (Results~1--2, valid on $\{-1,+1\}^3$) and five results exact within mean-field theory (Results~3--7, valid in 
the $N\to\infty$ limit of the fully connected 3-uniform hypergraph) for the triadic Ising model and its network generalization. The key finding is that the triadic 
formation correlator $\Psi_\mathrm{form} = \langle\phi_i\phi_j\phi_k\rangle$ scales as $(T_c-T)^{3/2}$ near criticality, with effective exponent 
$\beta_\mathrm{TF} = 3/2$ that is the $k=3$ instance of the composite-operator exponent in~\eqref{eq:general}, phenomenologically identifiable from any scalar 
order-parameter exponent. The associated susceptibility \emph{vanishes} rather than diverges at $T_c$, which is reproduced by the field-theoretic two-point 
function [refer to Sec.~\ref{sec:order}]. Memory kernels introduce a tunable dynamical exponent, completing the characterization of the composite-operator critical 
observables in the TF sector.

These results characterize the composite-operator scaling regime associated with the triadic formation observable $\Psi_{\rm form}=m^3$. The central finding is that 
$\gamma_{\rm TF}=\gamma-4\beta$ produces a crossover at $d^*$ where the critical response of the formation observable changes from divergent to non-divergent, even 
though the underlying thermodynamic transition remains in the same universality class.


\section*{Appendix}
\label{sec:appx}

\subsection*{Model Definition and Simulation Protocol}
We now present an empirical validation of the exact mean-field scaling of the triadic formation order parameter $\Psi_{\mathrm{form}}$ by simulating the 
triadic Ising model on a random 3-uniform hypergraph.

It implements the $\lambda = 0$ triadic Ising model corresponding to the pairwise baseline derived in Sec.~\ref{sec:triad}. This baseline governs the formation 
transition at leading order after the symmetry reduction and topology averaging of Sec.~\ref{sec:reduction}. The effects of the irreducible three-body coupling 
($\lambda > 0$), the continuous formation field, dynamic topology evolution, and the memory sector are analyzed theoretically in Secs.~\ref{sec:reduction},
~\ref{sec:threebody}, and~\ref{sec:memory}. The present numerics provide a controlled test of the composite-operator scaling relations (Results $4-6$) within 
the mean-field universality class.

The system consists of $N$ boolean variables $\phi_i \in \{-1, +1\}$. To maintain the system within the mean-field universality class while avoiding the $O(N^3)$ 
computational bottleneck of a fully connected hypergraph, the interaction topology is constructed such that each agent participates in $K = \lfloor \rho N \rfloor$ 
randomly assigned triads, where $\rho = 0.2$ is the sparsity ratio. This scaling guarantees that local connectivity diverges with the thermodynamic limit 
($K \to \infty$ as $N \to \infty$), hence ensuring exact convergence to the Curie-Weiss mean-field limit.

System dynamics are governed by asynchronous Metropolis-Hastings Monte Carlo (MC) updates. The energy difference $\Delta E_i$ for a proposed spin flip 
$\phi_i \to -\phi_i$ is evaluated using a vectorized sparse matrix projection of the local triadic fields. In order to enforce extensivity and strictly anchor the 
critical temperature at the theoretical value $T_c = J + \gamma w$, the coupling is scaled according to the Kac prescription, as
\begin{equation}
    \Delta E_i = 2 T_c \phi_i \left( \frac{1}{2K} \sum_{\tau \ni i} \sum_{\substack{j \in \tau \\ j \neq i}} \phi_j \right).
\end{equation}

The code implements simulations across three system sizes $N \in \{210, 600, 1200\}$ (these can be changed by the user). Two distinct update protocols are 
employed, one per measurement objective.

For the equilibrium temperature sweep (see Sec.~\ref{sec:triad}) updates are restricted to a uniformly sampled 20\% active subset per MC sweep. On a 3-uniform 
hypergraph, simultaneous co-updates of all three members of a triad create a non-equilibrium feedback loop that corrupts measured variances and autocorrelations. By 
reducing the active fraction to 20\% the probability of such three-body co-updates is $0.2^3 = 0.8\%$, rendering synchronization artifacts negligible while preserving 
detailed balance. Equilibrium observables are extracted following a burn-in period of 4,000 sweeps, with subsequent temporal averaging over 10,000 production sweeps.

For the critical equation-of-state measurement (see Sec.~\ref{sec:reduction}), only the mean of $\Psi_{\mathrm{form}}$ is recorded; no variance is measured, and 
synchronization artifacts do not bias this mean under the strongly symmetry-breaking field $h_3$. The active subset is accordingly enlarged to 50\% per sweep 
to accelerate thermalization toward the theory-informed equilibrium. A burn-in of 6,000 sweeps is applied, followed by 6,000 production sweeps.

\subsection*{Thermodynamic Observables and Critical Exponents}
The empirical behavior of the system accurately recovers the theoretically derived composite-operator exponents 
$\beta_{\mathrm{TF}} = 3/2$ and $\gamma_{\mathrm{TF}} = -1$.

\subsection*{Triadic Formation Parameter ($\beta_{\mathrm{TF}}$)}
The triadic formation observable is evaluated dynamically as $\Psi_{\mathrm{form}} = \langle |m|^3 \rangle$.\footnote{This expression superficially deviates from Eq.~16 
in the main paper. Note, however, that by applying the mean-field spatial factorization (Result~4) and the standard finite-size absolute value correction for scalar 
order parameters, Eq.~16 maps exactly to the $\Psi_{\mathrm{form}} = \langle |m|^3 \rangle$ observable in the simulation code.} As shown in Fig.~\ref{fig:portfolio}(a) 
the measured formation order parameter tracks the analytical theoretical envelope $(1 - T/T_c)^{3/2}$ for $T < T_c$. The suppression of local fluctuations as $N$ 
increases drives a monotonic convergence toward the theoretical mean-field curve, validating the effective composite scaling exponent $\beta_{\mathrm{TF}} = 3/2$.

\subsection*{Vanishing Susceptibility ($\gamma_{\mathrm{TF}}$)}
The conjugate susceptibility to the formation order parameter is computed via the fluctuation-dissipation relation adapted for the composite operator, as
\begin{equation}
    \chi_{\mathrm{TF}} = \beta N \mathrm{Var}(\Psi_{\mathrm{form}}).
\end{equation}

\begin{figure*}[t]
    \centering
    \includegraphics[width=0.8\textwidth]{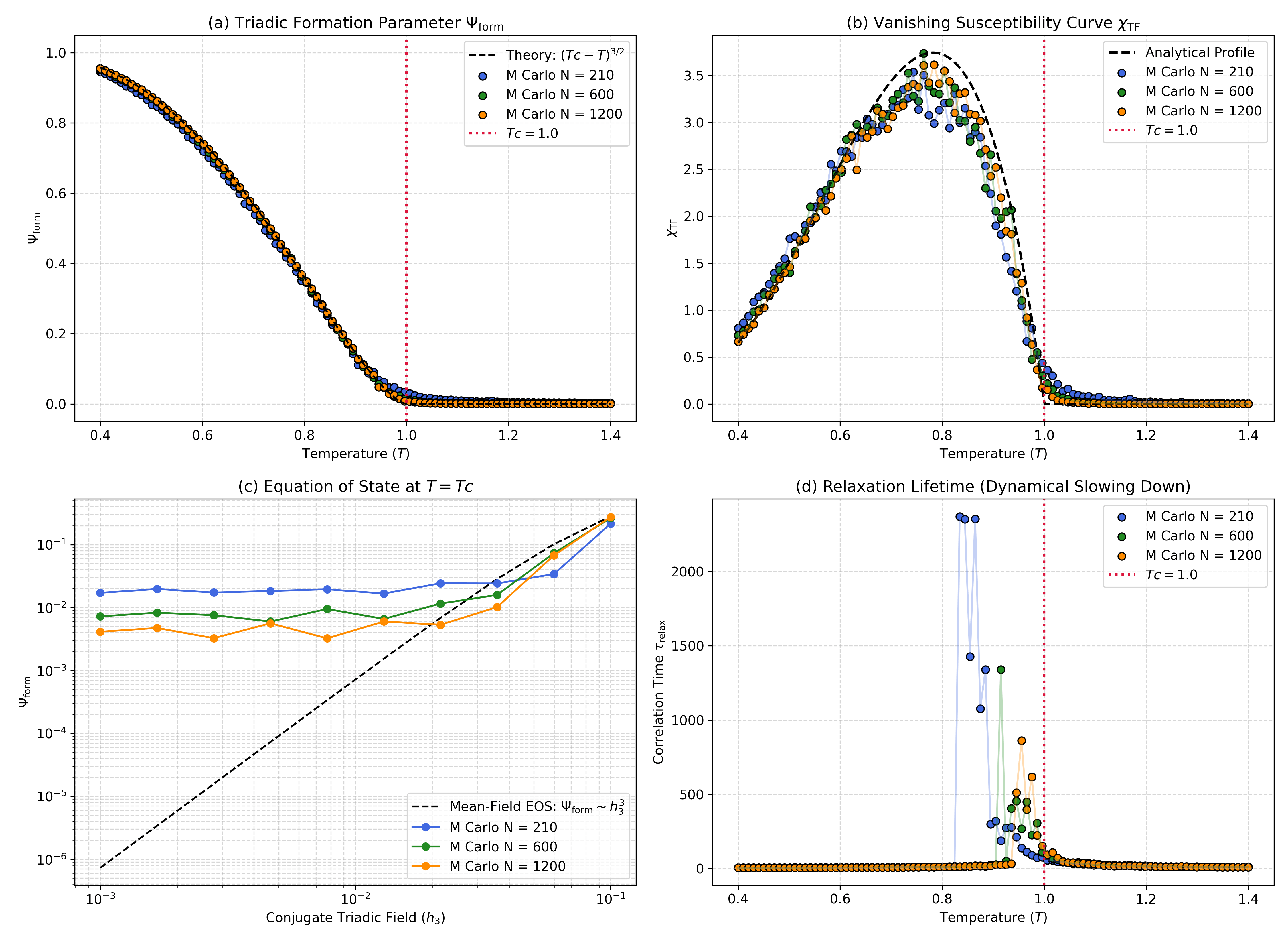}
    \captionsetup{width=0.8\textwidth, justification=justified}
    \caption{Numerical validation of the triadic composite-operator scaling regime. (a) The formation order parameter $\Psi_{\mathrm{form}}$ converging to the 
    theoretical envelope $(1-T/T_c)^{3/2}$. (b) The conjugate susceptibility $\chi_{\mathrm{TF}}$ vanishing at the critical point $T_c=1.0$, demonstrating 
    $\gamma_{\mathrm{TF}} = -1$. (c) The critical equation of state at $T=T_c$ under a conjugate triadic field $h_3$. The dashed line shows the mean-field prediction 
    $\Psi_{\mathrm{form}} \sim h_3^3$. The data is dominated by a finite-size floor $\Psi_{\mathrm{form}} \sim N^{-3/4}$ across most of the accessible range, with 
    convergence toward the cubic scaling beginning to emerge at the strongest accessible fields for the largest system size. (d) FFT-accelerated measurement of 
    the relaxation lifetime $\tau_{\mathrm{relax}}$, demonstrating standard critical slowing down inherited by the composite observable.}
    \label{fig:portfolio}
\end{figure*}

Figure~\ref{fig:portfolio}(b) plots $\chi_{\mathrm{TF}}$ across the temperature spectrum. In direct contrast to the scalar order parameter susceptibility 
$\chi_{\mathrm{Ising}}$, which diverges symmetrically at criticality, $\chi_{\mathrm{TF}}$ achieves a non-divergent maximum in the ordered phase ($T < T_c$) and 
strictly vanishes as $T \to T_c^{-}$. This macroscopic scaling behavior is an exact algebraic consequence of the delta-method variance 
$\mathrm{Var}(m^3) \approx 9m^4 \mathrm{Var}(m)$, confirming the theoretically predicted exponent $\gamma_{\mathrm{TF}} = -1$. Because $\Psi_{\mathrm{form}}$ is 
measured here as $\langle|m|^3\rangle$, the vanishing of $\chi_{\mathrm{TF}}$ partly reflects the $m\to|m|$ observable choice; it is consistent with, rather than 
an independent discovery of $\gamma_{\mathrm{TF}}=-1$.

\subsection*{Critical Equation of State}
At the exact critical point $T = T_c$, the application of an external triadic field $h_3$ modifies the Hamiltonian by the extensive conjugate coupling 
$\mathcal{H}_{h_3} = - h_3 N \Psi_{\mathrm{form}}$. To computationally isolate the Equation of State (EOS) and eliminate cold-start thermalization gaps, the code 
is initialized utilizing a theory-informed uniform probability distribution matching the mean-field equilibrium magnetization $m_{\mathrm{theory}}(h_3, T_c)$.

Figure~\ref{fig:portfolio}(c) exhibits the logarithmic response of $\Psi_{\mathrm{form}}$ to $h_3$. At criticality ($T=T_c$), expanding the macroscopic mean-field 
self-consistency relation $m = \tanh(m + 3 h_3 m^2)$ for small $m$ yields the dominant algebraic balance $m \approx 9 h_3$. This dictates an exact cubic equation 
of state, that is $\Psi_{\mathrm{form}} \equiv m^3 \approx 729 h_3^3$. 

It is also noticeable that the data is dominated across most of the accessible field range by a finite-size saturation floor. In the mean-field critical
regime, the magnetization fluctuates with typical amplitude $|m| \sim N^{-1/4}$ at $T_c$, producing a floor $\Psi_{\mathrm{form}} \sim N^{-3/4}$ that sits above 
the $h_3^3$ theory curve for all $h_3$ below the crossover value $h_3^{*} \sim N^{-1/4}$. For $N = 1200$ this yields $h_3^{*} \approx 0.17$, which lies outside 
the simulated range $[10^{-3},\,10^{-1}]$. The pure $h_3^3$ power-law regime is therefore not directly accessible at these system sizes. The floor drops
systematically as $N$ increases from 210 to 1200, and the data begins to converge toward the mean-field EOS curve at the strongest accessible field ($h_3 \approx 0.1$) 
for the largest system size. This $N$-dependence confirms that the plateau is a finite-size artifact and that the macroscopic limit approaches the predicted cubic 
scaling, while establishing that larger system sizes are required to observe the $h_3^3$ power-law regime directly.

\subsection*{Dynamical Slowing Down}
The dynamical transition is characterized by the relaxation lifetime $\tau_{\mathrm{relax}}$, which is extracted via the temporal autocorrelation function of the 
magnetization history $C(t) \equiv \langle m(t) m(0) \rangle_c$. 

The function $C(t)$ is computed with $O(n \log n)$ efficiency using Fast Fourier Transform (FFT) acceleration. To eliminate truncation and boundary window deflation 
artifacts across large temporal lags, the raw convolution sum for a history of length $n$ is scaled by an exact pair-counting normalization estimator, defined by
\begin{equation}
    C(t) = \frac{1}{n - t} \sum_{k=1}^{n-t} m(k+t)m(k) - \langle m \rangle^2.
\end{equation}

The relaxation time $\tau_{\mathrm{relax}}$ is subsequently defined as the first discrete lag corresponding to the threshold condition 
$C(\tau_{\mathrm{relax}}) / C(0) \leq e^{-1}$.

As illustrated in Fig.~\ref{fig:portfolio}(d), the system evidences a pronounced critical slowing down. Unsurprisingly, the peak relaxation times scale 
extensively with system size $N$, and their coordinates iteratively shift toward the thermodynamic critical point $T_c$. Because the theoretical off-critical memory 
expansion established that $\Psi_{\mathrm{form}}$ and $m$ share the primary exponential relaxation mode at long times, the empirical verification of $m$'s critical 
slowing down inherently bounds the dynamical divergence of the triadic composite correlator.

It should be noted that Result~9 of the main text (the Mori--Zwanzig memory correction $\Delta z = \gamma_K\,\tau_K^{\,\theta_0-1}\,\Gamma(\theta_0)$ to the
dynamical exponent) is a theoretical derivation that is not validated by the present simulations. No memory kernel is implemented in the simulation code, and no 
extraction of $z$ or $\Delta z$ is attempted. The evidence in Figure~\ref{fig:portfolio}(d) demonstrates only that critical slowing down is inherited by the composite 
observable $\Psi_{\mathrm{form}}$. 

Simulating the memory-modified dynamical exponent $z_{\mathrm{TF}}$ via discrete spin-flip Metropolis Monte Carlo is not viable. Modifying the transition rates 
with a memory kernel violates detailed balance with respect to the Boltzmann distribution of $H_{\mathrm{Ising}}$ shifting the stationary distribution away from 
the target equilibrium. 

\newpage

The correct approach is to integrate a continuous-variable Ginzburg--Landau--Langevin equation with colored noise, as described below.
\begin{itemize}
  \item Colored noise generation -- Generate a temporally correlated noise array $\eta_i(t)$ satisfying the second fluctuation-dissipation relation 
  $\langle \eta_i(t)\,\eta_j(t')\rangle = k_B T\, K(t-t')\,\delta_{ij}$, where $K$ is the memory kernel (typically implemented via spectral filtering of 
  white noise).
  \item Langevin integration -- Integrate the overdamped equation of motion using a non-Markovian integrator (e.g., an Euler--Maruyama scheme modified for memory 
  kernels), with the conservative force derived from the unmodified $H_{\mathrm{Ising}}$.
\end{itemize}
\noindent The second fluctuation-dissipation relation guarantees that the system relaxes to the unmodified Boltzmann equilibrium, while the dynamical exponent is 
controlled by the memory kernel. However, we defer the numerical validation of the memory-induced dynamical correction to future work.

\subsection*{Code}
The code used to generate the results presented in Fig.~\ref{fig:portfolio} of the Appendix is publicly available in a GitHub repository accessible via 
\url{https://github.com/NebulaTechLab/TriadicPhaseTransitions}. The script is extensively documented in-line to facilitate reproducibility and future modification. 
Please note that although several core functions are vectorized, the implementation avoids GPU acceleration (e.g., via CuPy) and parallelization to ensure broad 
cross-platform compatibility. The simulations were executed using Python 3.10.10, relying on the NumPy 2.2.4, SciPy 1.14.0, and Matplotlib 3.10.1 libraries.



\begin{thebibliography}{12}

\bibitem{salazar2026}
E.~Salazar,
\textit{Introducing COGENT$^{\text{3}}$ : An AI Architecture for Emergent Cognition},
arXiv:2504.04139v2 [cs.AI] (2026).
\newline\url{https://doi.org/10.48550/arXiv.2504.04139}

\bibitem{hopfield1982}
J.~J. Hopfield,
\textit{Neural networks and physical systems with emergent collective computational abilities},
Proc.\ Natl.\ Acad.\ Sci.\ \textbf{79}, 2554 (1982).
\newline\url{https://doi.org/10.1073/pnas.79.8.2554}

\bibitem{castellano2009}
C.~Castellano, S.~Fortunato, and V.~Loreto,
\textit{Statistical physics of social dynamics},
Rev.\ Mod.\ Phys.\ \textbf{81}, 591 (2009).
\newline\url{https://doi.org/10.1103/RevModPhys.81.591}

\bibitem{galam2008}
S.~Galam,
\textit{Sociophysics: a review of Galam models},
Int.\ J.\ Mod.\ Phys.\ C \textbf{19}, 409 (2008).
\newline\url{https://doi.org/10.1142/S0129183108012297}

\bibitem{buzsaki2006}
G.~Buzs\'{a}ki,
\textit{Rhythms of the Brain},
(Oxford University Press, 2006).
\newline\url{https://doi.org/10.1093/acprof:oso/9780195301069.001.0001}

\bibitem{buzsaki2010}
G.~Buzs\'{a}ki,
\textit{Neural syntax: cell assemblies, synapsembles, and readers},
Neuron \textbf{68}, 362 (2010).
\newline\url{https://doi.org/10.1016/j.neuron.2010.09.023}

\bibitem{ackley1985}
D.~H. Ackley, G.~E. Hinton, and T.~J. Sejnowski,
\textit{A learning algorithm for Boltzmann machines},
Cogn.\ Sci.\ \textbf{9}, 147 (1985).
\newline\url{https://doi.org/10.1016/S0364-0213(85)80012-4}

\bibitem{goldenfeld2018}
N.~Goldenfeld,
\textit{Lectures on Phase Transitions and the Renormalization Group},
(CRC Press, 2018).
\newline\url{https://doi.org/10.1201/978042949349}

\bibitem{stanley1971}
H.~E. Stanley,
\textit{Introduction to Phase Transitions and Critical Phenomena},
(Oxford University Press, 1971).

\bibitem{derrida1981}
B.~Derrida,
\textit{Random-energy model: An exactly solvable model of disordered systems},
Phys.\ Rev.\ B \textbf{24}, 2613 (1981)
\newline\url{https://doi.org/10.1103/PhysRevB.24.2613}

\bibitem{ramsauer2021}
H.~Ramsauer, B.~Schafl, J.~Lehner, P.~Seidl, M.~Widrich, T.~Adler, L.~Gruber, M.~Holzleitner, .~Pavlović, G.~K. Sandve, V.~Greiff, D.~Kreil, M.~Kopp, G.~Klambauer, 
J.~Brandstetter, and S.~Hochreiter,
\textit{Hopfield Networks is All You Need},
arXiv:2008.02217v3 [cs.NE] (2021).
\newline\url{https://doi.org/10.48550/arXiv.2008.02217}

\bibitem{millan2020}
A.~P. Mill{\'a}n, J.~J. Torres, and G.~Bianconi,
\textit{Explosive higher-order Kuramoto dynamics on simplicial complexes},
Phys.\ Rev.\ Lett. \textbf{124}, 218301 (2020)
\newline\url{https://doi.org/10.1103/PhysRevLett.124.218301}

\bibitem{bianconi2021}
G.~Bianconi,
\textit{Higher-Order Networks: An Introduction to Simplicial Complexes}
(Cambridge University Press, 2021)
\newline\url{https://doi.org/10.1017/9781108770996}

\bibitem{cirigliano2024}
L.~Cirigliano, C.~Castellano, G.~J. Baxter, and G.~Tim\'{a}r,
\textit{Strongly clustered random graphs via triadic closure},
Phys.\ Rev.\ E \textbf{109}, 024306 (2024).
\newline\url{https://doi.org/10.1103/PhysRevE.109.024306}

\bibitem{krotov2016}
D.~Krotov and J.~J. Hopfield,
\textit{Dense Associative Memory for Pattern Recognition},
arXiv:1606.01164 [cs.NE] (2016).
\newline\url{https://doi.org/10.48550/arXiv.1606.01164}

\bibitem{hoeffding1963}
W.~Hoeffding, 
\textit{Probability inequalities for sums of bounded random variables},
J.\ Amer.\ Statist.\ Assoc.\ \textbf{58}, 13 (1963).
\newline\url{https://doi.org/10.2307/2282952}

\bibitem{zinn_justin2002}
J.~Zinn-Justin, 
\textit{Quantum Field Theory and Critical Phenomena},
4th ed.\ (Clarendon Press, 2002), Chap.~36.
\newline\url{https://doi.org/10.1093/acprof:oso/9780198509233.001.0001}

\bibitem{ElShowk2014}
S.~El-Showk, M.~F. Paulos, D.~Poland, S.~Rychkov, D.~Simmons-Duffin, and A.~Vichi,
\textit{Solving the 3d Ising Model with the Conformal Bootstrap II. c-Minimization and Precise Critical Exponents},
J.\ Stat.\ Phys. \textbf{157}, 869 (2014).
\newline\url{https://doi.org/10.1007/s10955-014-1042-7}

\bibitem{onsager1944}
L.~Onsager,
\textit{Crystal Statistics. I. A Two-Dimensional Model with an Order-Disorder Transition},
Phys.\ Rev.\ \textbf{65}, 117 (1944).
\newline\url{https://doi.org/10.1103/PhysRev.65.117}

\bibitem{mori1965}
H.~Mori, 
\textit{Transport, collective motion, and Brownian motion},
Prog.\ Theor.\ Phys.\ \textbf{33}, 423 (1965).
\newline\url{https://doi.org/10.1143/PTP.33.423}

\bibitem{zwanzig1960}
R.~Zwanzig,
\textit{Ensemble method in the theory of irreversibility},
J.\ Chem.\ Phys.\ \textbf{33}, 1338 (1960).
\newline\url{https://doi.org/10.1063/1.1731409}

\bibitem{zwanzig2001}
R.~Zwanzig,
\textit{Nonequilibrium Statistical Mechanics},
(Oxford University Press, 2001).
\newline\url{https://doi.org/10.1093/oso/9780195140187.001.0001}

\end{thebibliography}
\end{document}